\documentclass[aps,reprint,amsmath,amssymb,graphicx]{revtex4-2}
\usepackage[utf8]{inputenc}
\usepackage{natbib}
\usepackage{color}
\usepackage{xcolor}
\usepackage{graphicx}
\usepackage{mdframed}
\usepackage{physics}
\usepackage{bm}
\begin{document}


\title{Neutral Theory for competing attention in social networks}
\author{Carlos A. Plata$^1$, Emanuele Pigani$^{1}$, Sandro Azaele$^1$,Violeta Calleja-Solanas$^2$, Mar\'ia J.~Palazzi$^3$, Albert Sol\'e-Ribalta$^3$,  Javier Borge-Holthoefer$^3$, Sandro Meloni$^2$, Samir Suweis$^1$ \\}
\address{$^1$Dipartimento di Fisica ``G. Galilei'', Università di Padova,
Via Marzolo 8, 35131 Padova, Italy \\
$^2$IFISC, Institute for Cross-Disciplinary Physics and Complex Systems (CSIC-UIB), 07122, Palma de Mallorca, Spain\\
$^3$Internet Interdisciplinary Institute (IN3), Universitat Oberta de Catalunya, Barcelona, Catalonia, Spain}



\begin{abstract}
Online social networks have significantly changed global communication dynamics. Although the complexity of such dynamics, emergent patterns from empirical micro-blogging data on hashtag usage, such as (but not limited to) meme popularity power law distributions have been observed. However, going beyond empirical or simulation-based studies, to highlight and distinguish fundamental mechanisms at the origin of such patterns, is still an open problem.
Herein, inspired by Neutral Theory of ecology, we provide an analytical formalism capable of capturing the competition for attention in online social systems and explaining  several observed emergent patterns in the context of information ecosystems. Specifically, besides meme popularity distributions, we study the persistence distributions of memes within the network and the capacity of the system to sustain several coexisting diverse memes. Such properties have clear analogy with extinction time and biodiversity in natural ecosystems. In the social network context, diversity of ideas is indeed a fingerprint of the ``health'' of the communication process. We finally show how our theoretical framework is successful in explaining real data collected  from the Twitter online social network. Henceforth, the proposed framework  introduced here brings both, new insights to the role played by competition and network structure on memes diversity and persistence, and an applicable null model to be tested against real data. 
\end{abstract}

\maketitle

\section{\label{sec:introduction}Introduction}

An online social network (OSN) is a virtual social structure made of individuals using the Internet as a communication medium for interacting, sharing contents and opinions.
OSNs allow hundreds of millions of Internet users worldwide to produce and consume contents, providing access to a very vast source of information on an unprecedented scale. 
Therefore, nowadays OSNs constitute mainstream communication channels to interact, exchange opinions, and reach consensus.

In recent years, it has increasingly become evident that competition significantly shapes the structure and the dynamics on these information-driven platforms \cite{weng2012competition,altman2013competition,kleineberg2016competition}: users thrive for visibility, while memes  can be thought of as entities that compete for users' attention.

Nevertheless, it is hard to disentangle the effects of limited attention from many concurrent factors \cite{romero2011differences}, such as the structure of the underlying social network \cite{centola2010spread, borge2011structural}, the activity of users \cite{lerman2010information}, the different degrees of influence of information spreaders \cite{romero2011influence}, the intrinsic quality of the information they spread \cite{qiu2017limited}, and the persistence of topics \cite{asur2011trends}.

Through the analysis of OSNs data, emergent properties, such as the emergence of consensus \cite{weng2012competition}, viral spreading \cite{weng2013virality}, power-law distribution of memes popularity \cite{gleeson2016effects}  and echo-chambers effects \cite{del2016spreading,baumann2020modeling}, have been observed even among different platforms. Nevertheless, the availability of massive streams of online data does not give per se theoretical insights to understand these complex inter-plays among the structure of OSNs, memes popularity and users attention dynamics.

The aforementioned  ubiquitous properties are signatures of the emergent simplicity characterizing complex systems. Therefore, it  naturally calls for a statistical mechanics approach, i.e., the attempt to understand regularities at large scale as collective effects of the dynamics at the individual/meme scale. 
The bridge between these scales  would also allow to better understand how these macroscopic (system-wide) patterns are affected by the microscopic dynamics of the interacting elements (users and memes) forming the OSN.

One of the first attempts to unveil the emergence of fat-tailed power-law distributions for memes popularity starting from an interacting particle model is the work of Gleeson and collaborators  \cite{gleeson2014competition}. Therein, they studied a stochastic model with simple microscopic rules to describe the evolution, that is, the spreading of memes competing for the limited resource of users attention, on a Twitter-like OSN. In real life, each user in the OSN pays attention to a finite number of memes constrained by the finite capacity of users. In the model, this picture is simplified assuming that each user can just pay attention to a single meme; although further generalizations have already been considered \cite{gleeson2016effects}. 

The type of stochastic processes introduced in \cite{gleeson2014competition} has been used for decades in population genetics \cite{kimura1983neutral} and in ecology \cite{hubbell2001unified}. In particular, these models, based on simple stochastic rules, which neglect species (genes) fitness, represent ``null'' neutral models in which no intrinsic advantage is ascribed to a particular type of species (genes). Therefore, the fate of a species depends on its topological role in the network  and on random demographic effects.  

The concept of neutrality and the so-called neutral models have attracted a lot of interest in the communities of statistical physicists \cite{vallade2003,blythe2007stochastic,martinello2017}. Indeed, neutral theory has been proven to be very successful when describing ubiquitous emergent patterns, which do not depend on the specific details of the system under study \cite{Volkov2003,Volkov2007,bertuzzo2011spatial,azaele2016neutral}. 

The analogy between memes competing for attention in an OSN and species competing for resources in an ecosystem suggests that the approaches introduced in the framework of ecological neutral theories can be exploited to extract novel relevant information about the dynamics of users attention. Different memes can be thought of as different species and, in addition to the spreading dynamics, new memes are introduced in the system by innovations events (speciation processes).

Such qualitative similarities between natural and ``information'' ecosystems have already been explored towards more quantitative accounts. Along this direction, Borge-Holthoefer et al. \cite{borge2017emergence} have found evidence of nested structural signatures --a landmark feature in natural mutualistic assemblages \cite{bascompte2003nested,suweis2013emergence,suweis2015effect,payrato2019breaking} -- when analyzing time-resolved online communication discussions after external information ``shocks'', e.g. breaking news. 
Similarly, Lorenz and collaborators \cite{lorenz2019accelerating}, using a mathematical model based on Lotka-Volterra equations, have been able to explain some empirical data patterns in different OSNs. Moreover, they have suggested that relatively fast loss of attention is driven by increasing production and consumption of contents, leading to higher turnover rates and shorter collective attention for individual topics.

Following on this line of research, here we propose an analytically tractable neutral theory for competing attention in OSNs. In particular, we derive the equations for the evolution of users' attention to the different memes. The approach starts from a Master Equation describing the stochastic evolution of users' attention to a meme. In the limit of large networks, the equation is well approximated by a Fokker-Planck equation which resembles the dynamics of species abundances in ecological neutral theories \cite{azaele2006}. We thus analytically compute several new quantities of interest to characterize communication dynamics such as: the number of different coexisting memes - a measure of diversity of coexisting ideas in the OSN -, the distribution of user attention to these memes, and the average persistence time of user's attention on a meme. Moreover, to investigate how the emergent properties depend on the underlying network of users interactions, we compare our theoretical predictions to numerical simulations of the model, and finally, test our approach against empirical stream-data from the Twitter OSN .  

The work is organized as follows. In the next section, we introduce the interacting particle model describing memes propagation. In section \ref{sec:results} we derive the evolution equation for competing attention in OSNs. Therein, we present both (exact) discrete and (approximated) continuous description of the system, computing several relevant patterns of the OSN related to attention. As expected, the evolution of users attention depends on the OSN structure. Moreover, therein we compare --successfully validating-- the results of the model with real data collected from Twitter. Finally, in section \ref{sec:conclusions}, we summarize and discuss the achievements of the proposed neutral model to the study of competing attentions in social networks.

\section{A Neutral Model for user attention}
\label{sec:model}

Let us consider a directed OSN of $N$ nodes where each of them represents a user. The network is solely characterized by its out-degree distribution $p_k$, that is, a random user has $k$ followers with probability $p_k$. Each node can be thought of as the screen of the individual displaying the meme of current interest for that user. For the sake of simplicity, we assume that each screen has capacity for only one meme although some generalizations to this respect are possible \cite{gleeson2014competition,gleeson2016effects}. Therefore, the state of the system at time $\tilde{t}$ is given by the list of memes appearing in all the nodes.

\begin{figure}
   \includegraphics[width=0.49\textwidth]{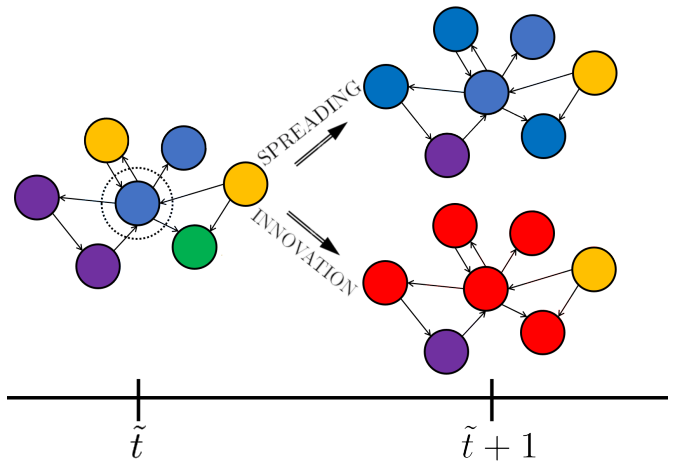}
    \caption{Sketch of the model. Each node represents a user in the network with the color denoting the different memes on its screen. The directed edges stand for the relations of following. Specifically, our convention is that the edge starts in the node which is followed and ends in the follower. The node chosen to spread the meme is highlighted with a dotted circle. A time step is represented with its two possible outcomes, either an innovation event (with probability $\tilde{\mu}$) or a spreading event (with probability $1-\tilde{\mu}$).}
   \label{Fig1}
\end{figure}

The dynamics is introduced in discrete time steps representing the subsequent times where an action is carried out by any of the users. During each time step one node is picked at random and it either (\emph{i}) (re)tweets the meme currently on its screen to all its followers with probability $1-\tilde{\mu}$, while its own screen remains unchanged; or (\emph{ii}) innovates with probability $\tilde{\mu}$ generating a brand-new meme that appears on its screen and is tweeted to all the followers. An illustrative visualization of one time step of the dynamics is shown in Fig.~\ref{Fig1}.

The configurational state of the system in a given time is described by the correspondence list of memes the users are paying attention to. We define the set of \textit{attention} variables  $x_m$ as the fraction of nodes in the system which are paying attention to meme $m$. Consequently, the normalization constraint $\sum_m x_m =1$ holds for all time.

Another relevant variable in the simulated OSN, as originally introduced by Gleeson and collaborators \cite{gleeson2014competition} is the memes \textit{popularity}. The popularity $n_m$ of meme $m$ is defined as the number of times that the meme $m$ has been broadcasted. Because of this definition, $n_m$ is a non-decreasing function of time, which increases by one every time $m$ is (re)tweeted. On the contrary, the evolution of attention $x_m$ can either increase or decrease until the meme disappears from the system. 
Note that, in our model, once a meme disappears from the OSN, there is no mechanism to re-introduce it in the system (innovation brings always brand-new memes). Therefore, once a meme goes extinct, its popularity remains constant for the remaining simulation time.

In the aforementioned work by Gleeson et al.~\cite{gleeson2014competition}, they semi-analytically find, and compare to numeric simulations, asymptotic power-laws behavior for the popularity distribution. In particular they have investigated the cases where $p_k=\delta_{k,k^*}$ (constant number of followers for all users), and $p_k \propto k^{-\gamma}$, i.e. a power-law distribution for the number of followers. 
In both cases, they have found that memes popularity has a fat tail distribution, but with different exponents \cite{gleeson2014competition}.
Popularity can be measured from real Twitter data streams, as the number of (re)tweets since a given hashtag appears in the OSN. Although a systematic analysis of memes popularity distributions is still lacking,
there are some results suggesting that indeed they display asymptotic power-law behavior \cite{gleeson2016effects}, pointing out thus that the proposed model is a good candidate to elucidate at least this pattern observed in empirical OSNs. In the next section, we show how in fact, exploiting neutral theory, we can go beyond meme popularity distribution and analytically study also the diversity of coexisting ideas in the OSN and the the distribution of users attention to such coexisting memes, and also the average persistence time of user's attention on a meme.


 


\section{Neutral emergent patterns in users attention}
\label{sec:results}


Because of the neutrality hypothesis, we assume complete symmetry among memes (i.e., we neglect meme identities or labels) and we thus consider the system as an effective two-memes system with large but finite $N$. We call $A$ the meme we are focusing on, whereas $B$ simply represents the collection of memes that are non-$A$. The variable $x \equiv x_A$ describes the fraction of users that are paying attention to meme $A$ (i.e., the meme is on their screen) and hence $x_B=1-x$. It comes in handy to define the total number of users paying attention to meme $A$ as $\nu$, i.e., $\nu \equiv x\cdot N$.

The dynamics, as described in the previous section, is defined in discrete time steps and obeys the Markovian property, i.e. the transition probability to the next state depends only on the current state. Consequently, we can properly formulate the time evolution of the probability $P\left(\nu,\tilde{t}\right)$ of having $\nu$ users paying attention to meme $A$ at the time step  $\tilde{t}$ by means of a discrete Master Equation, which reads

\begin{align}
\label{eq:ME}
P\left(\nu,\tilde{t}+1\right)&-P\left(\nu,\tilde{t}\right)= \sum_{k=0}^{N-1}  \sum_{j=0}^{k} \bigg\{W_{k,j}^{A,s}\left( \nu - j \right)  P \left( \nu - j ,\tilde{t}\right) 
\nonumber \\
 & + W_{k,j}^{A,i}\left( \nu + j +1 \right)  P \left( \nu + j +1,\tilde{t}\right)
\nonumber \\
 &+W_{k,j}^{B}\left( \nu + j \right)  P \left( \nu + j ,\tilde{t}\right) 
 \nonumber \\
 &-  \left[ W_{k,j}^{A,s}\left( \nu \right)+W_{k,j}^{A,i}\left( \nu \right)+W_{k,j}^{B}\left( \nu \right)\right] P \left( \nu  ,\tilde{t}\right) \bigg\}.
\end{align}

In the Master Equation \eqref{eq:ME}, we have taken into account that there are three main different classes of transformations with different transition probabilities $W$. Namely, the selected node to act in a particular time step may be either $A$ or $B$: this is indicated by the the first superscript in the transition probabilities $W$. The second superscript denotes the type of event, i.e. $\{s\}$ ~if the selected node spreads the current meme with probability $1-\tilde{\mu}$ or $\{i\}$ if it first innovates with probability $\tilde{\mu}$ and then spread the new meme. Note that, since $B$ is defined as non-$A$, innovation of $B$ generates also a $B$ meme and thus there is no need to differentiate between spread or innovation when the selected node contains a $B$ meme. The subscripts give information about the connected nodes of the selected node (its \emph{followers} ). We define $k \in \{0, \ldots ,N-1\}$ as  the number of the followers of the selected node, whereas $j\in \{0, \ldots,k\}$ is the
number of followers having a meme on their screen that is different from the meme spreading from the $i$ node. 
Thus, the dependence on the properties of the network is carried in the specific form of the transition rates. The detailed form of the transition rates $W$ and the physical meaning  of the Master Equation \eqref{eq:ME} are provided in Appendix \ref{app:FP}.

Assuming the innovation rate scales as $\tilde{\mu}=\mu/N$ and defining the continuous time variable $t=\tilde{t}/N^2$, it is possible to perform diffusive approximation of the Master Equation \eqref{eq:ME}. The Kramers-Moyal expansion \cite{vKampen,azaele2016neutral} yields the Fokker-Planck equation 
\begin{equation}
\label{eq:FP}
\partial_t P(x,t) = - \partial_x \left[ A(x) P(x,t) \right] + \frac{1}{2} \partial_x^2 \left[ B(x) P(x,t) \right], 
\end{equation}
with respectively drift and diffusion coefficients
\begin{subequations}
\label{eq:FP-coef}
\begin{align}
&A(x)=-a x, &a=\mu \left( 1+\left\langle k \right\rangle \right), \label{eq:FP-coef1} \\
&B(x)=bx(1-x), &b=  \left( 1+\left\langle k \right\rangle \right)\left\langle k \right\rangle + \sigma^2_k. \label{eq:FP-coef2}
\end{align}
\end{subequations}
The drift term is negative and drives the memes to extinction with a rate that increases with the innovation rate and the mean number of followers in the network $\left\langle k \right\rangle=\sum_k kp_k$. The dependence of the diffusion term on $x(1-x)$ \cite{al2005langevin} indicates that there is a limit (a carrying capacity, in the language of ecology) in the attention that both competing memes can collect, as we are considering a finite number of users in the system. The constant $b$ increases the (demographic) fluctuations with $\left\langle k \right\rangle$ and the variability of the network through the variance $\sigma^2_k=\sum_k \left(k-\left\langle k \right\rangle \right)^2 p_k $. Detailed derivation of the Fokker-Planck equation \eqref{eq:FP} associated to the Master Equation \eqref{eq:ME} is provided in Appendix \ref{app:FP}.

Thanks to the Master Equation \eqref{eq:ME} and the Fokker-Planck equation \eqref{eq:FP}, it is possible to study the emergent patterns in users' attention in the proposed OSN neutral model.

\subsection{Mean persistence time for attention} 

The dynamics of the users' attention has an absorbing boundary at  $x=0$, i.e. the attention to a given meme will eventually go to zero. 
The time that a meme persists in the OSN receiving attention is a relevant quantity to study to better understand communication dynamics, as it gives insights on the virality of that meme, i.e., the longer it remains ``active'' the more probable it may go viral. In the ecological context, species persistence times \cite{bertuzzo2011spatial} or lifetimes \cite{pigolotti2005} have been thoroughly studied in different ecological neutral models and proved to be able to describe and thus elucidate the ecological mechanisms behind the power-law shape (with exponential cut-off) of species persistence patterns observed in real ecosystems \cite{suweis2012species}. Due to the stochastic nature of the neutral dynamics of the memes, these persistence times are random variables identically distributed among memes. Herein, we focus on the computation of the mean persistence time of a meme and its variance.

Previously, we have derived the forward evolution equations that govern the probability density function, $P(x,t)$, of the attention. The probability density function $Q(x_0,t)$ of reaching the absorbing boundary at zero at time $t$, departing from initial attention $x_0$,  obeys the backward Fokker-Planck equation \cite{RednerBook}
\begin{equation}
\label{eq:back-FP}
\partial_t Q(x_0,t) = A(x_0) \partial_{x_0}   Q(x_0,t)  + \frac{1}{2} B(x_0) \partial_{x_0}^2   Q(x_0,t),
\end{equation}
with the same $A$ and $B$ defined in \eqref{eq:FP-coef}.

Multiplying \eqref{eq:back-FP} by $t$ and integrating over all times, one obtains the differential equation that governs the mean persistence time $\tau(x_0)$ to reach zero departing from $x_0$. That equation can be analytically solved with the proper boundary conditions (see  Appendix \ref{app:FP-back} for details) giving the result
\begin{align}
\label{eq:tau_FP}
\tau(x_0) =&- \Bigg\{(1-x_0)\,\, {}_2F_1\left(1, \frac{2a}{b};1 + \frac{2a}{b} ;1-x_0 \right) 
\nonumber \\
& \qquad +  \frac{2a}{b} \left[ \gamma + \psi \left( \frac{2a}{b} \right) + \ln x_0 \right]\Bigg\} 
\nonumber \\
 &\times \left[ a \left(1-\frac{2a}{b} \right) \right]^{-1},
\end{align} 
where $_2F_1$ stands for the ordinary hypergeometric function,  $\gamma$ is the Euler--Mascheroni constant, $\psi$ denotes the digamma function, and the pair $\{a,b\}$ is defined in \eqref{eq:FP-coef}. 

Having an analytical form for $\tau(x_0)$ allows us to study how the persistence of attention varies for the OSN's innovation rate, and structure. We find that $\tau$ is a monotonic decreasing function of $\mu$, $\left\langle k \right\rangle$ and $\sigma_k^2$, taking the rest of parameters fixed.  

We find that, as expected, higher innovation rate favors the demise of living memes, and thus decreases the average persistence time on meme's attention. Similarly, higher connectivity of the OSN, measured by the average network degree, also drives, \textit{on average}, to a faster decay of attention on existing memes. This result can be understood by observing that large $\langle k \rangle$ increases the diffusivity of the memes spreading in the systems, thus accelerating the time at which the attention to the meme will reach zero. Finally, we find that for fixed average connectivity, heterogeneity of the network, as measured by $\sigma_k^2$, disfavours, \textit{on average},  long persistent attention on a given meme. In other words, we find that the presence of large hubs (\emph{influencers}) coexisting with many poorly connected nodes (\emph{users}) drives the communication dynamics towards a state dominated by fast, short persistent tweets, which disappear relatively quickly from the OSN.

Note that \eqref{eq:tau_FP} is an approximated result, since it comes out from the diffusive continuum limit. To go beyond this approximation, we write the backward Master Equation for the discrete system, describing the evolution of the probability  $Q(\nu_0,\tilde{t})$ for a meme to reach zero users attention at time $\tilde{t}$, if at the initial time the meme has $\nu_0$ users following it. This equation reads
\begin{align}
\label{eq:back-ME}
Q\left(\nu_0,\tilde{t}+1\right)&-Q\left(\nu_0,\tilde{t}\right)= \sum_{k=0}^{N-1}  \sum_{j=0}^{k}  \Bigg\{W_{k,j}^{A,s}\left( \nu_0  \right)  Q \left( \nu_0 + j ,\tilde{t}\right)  \nonumber \\& +W_{k,j}^{A,i}\left( \nu_0  \right)  Q \left( \nu_0 - j -1,\tilde{t}\right) 
\nonumber
\\
& +W_{k,j}^{B}\left( \nu_0 \right)  Q \left( \nu_0 - j ,\tilde{t}\right)  \Bigg\} - Q \left( \nu_0  ,\tilde{t}\right) .
\end{align}

Multiplying by $\tilde{t}$ the above backward Master Equation and summing over all times, we obtain a system of linear equations whose solution is the vector of mean persistence times $\tilde{\bm{\tau}}$, whose components are $\tilde{\tau}(\nu_0)$ with $\nu_0 \in \{ 1,\ldots,N \}$. This equation, written in matrix form, yields
\begin{equation}
\label{eq:tau-exact0}
\tilde{\bm{\tau}} = \mathbb{M} \tilde{\bm{\tau}} + \bm{1},
\end{equation}
where the components of the matrix $\mathbb{M}$ are appropriate combinations of the transition probabilities $W$ (see Appendix \ref{app:ME-back} for the details) and $\bm{1}$ is a $N$-dimensional vector of ones. Therefore, the exact result for the mean persistence times is given by
\begin{equation}
\label{eq:tau-exact}
\tilde{\bm{\tau}}= (\mathbb{I}-\mathbb{M})^{-1}\bm{1},
\end{equation} 
with $\mathbb{I}$ being the identity matrix in dimension $N$. 
Furthermore, one can obtain higher order moments using \eqref{eq:back-ME}, multiplying it by powers of $\tilde{t}$ and summing over all times. Specifically, we have done so in order to obtain the meme attention variance $\sigma^2_{\tilde{\tau}}= \langle \tilde{t}^2 \rangle - \tilde{\tau}^2$. The calculation of the  mean squared attention persistence time results
\begin{equation}
\label{eq:tau2-exact}
\langle \tilde{\bm{t}}^2 \rangle = (\mathbb{I}-\mathbb{M})^{-1} (2 \tilde{\bm{\tau}}-\bm{1}).
\end{equation} 
Detailed calculations are provided in Appendix \ref{app:ME-back}.

In Fig.~\ref{Fig2} we have compared the results  obtained from numerical simulation of the model regarding persistence time of memes with our theoretical predictions. We have done simulations for a system with  $N=10^3$ users.  The two OSN structures chosen along this work in order to study the role of the network structure are: (i) homogeneous networks, i.e., $p_k=\delta_{k,k^*}$ with constant $k^*$ and thus all users have the same number of followers, and (ii) scale-free networks with $p_k \propto k^{-\alpha}$ for $k \geq k_{min}$.  For further details on the simulation algorithm, see Appendix \ref{app:sim}.

The value $k^*=10$ has been chosen so to avoid falling in a regime very far from the diffusive approximation. As seen in the panel (a) of Fig.~\ref{Fig2}, the exact results perfectly agree with the simulations for both mean value $\tau$ and standard deviation $\sigma_{\tau}$. Remarkably, in spite of the diffusive approximation, the prediction given by \eqref{eq:tau_FP} still gives a very good estimation of the persistence time of the memes. This relative success of the diffusive approximation is reasonable in a system where there are not big chances of having burst events. Nevertheless, we expect the diffusive theory to fail when the distribution allows such burst events. In fact, if $k^*$ is very large, then in a single step it is likely that most of the users in the network change their attention, and thus the dynamics regime for the attention is bursty rather than diffusive. This is the case for example in scale free OSNs, where $p_k$ presents long tails, i.e. few nodes have very large degree centrality. 

Indeed, the most interesting case is when users have a distribution that is strongly uneven as in real OSNs. To investigate the impact of such heterogeneous structure on the dynamics, we have considered  a power-law out-degree distribution which has been also found empirically \cite{gleeson2016effects} in some OSNs. More specifically, we assume that $p_k$ is zero for $k<k_{min}$ and  $p_k \propto k^{-\alpha}$ for $k \geq k_{min}$ \cite{gleeson2014competition}. In particular, in the following we consider $k_{min}=4$ and $\alpha=2.5$ so that we have an average number of followers $\left\langle k \right\rangle \approx 10$ close to the previous case $k^*=10$, but with much higher variance. In the panel (b) of Fig.~\ref{Fig2} we observe clearly that the result of the exact approach using the backward Master Equation matches accurately the values found in the numerical simulations. The diffusive approximation, instead, as expected because of the presence of bursting events, fails to quantitatively describe the observed patterns.

\begin{figure*}
   \includegraphics[width=0.49\textwidth]{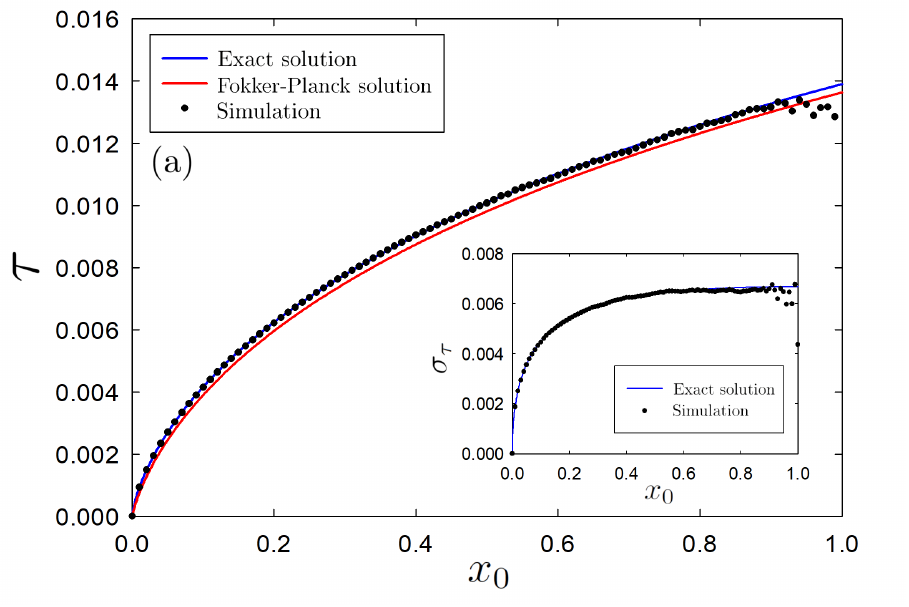}
   \includegraphics[width=0.49\textwidth]{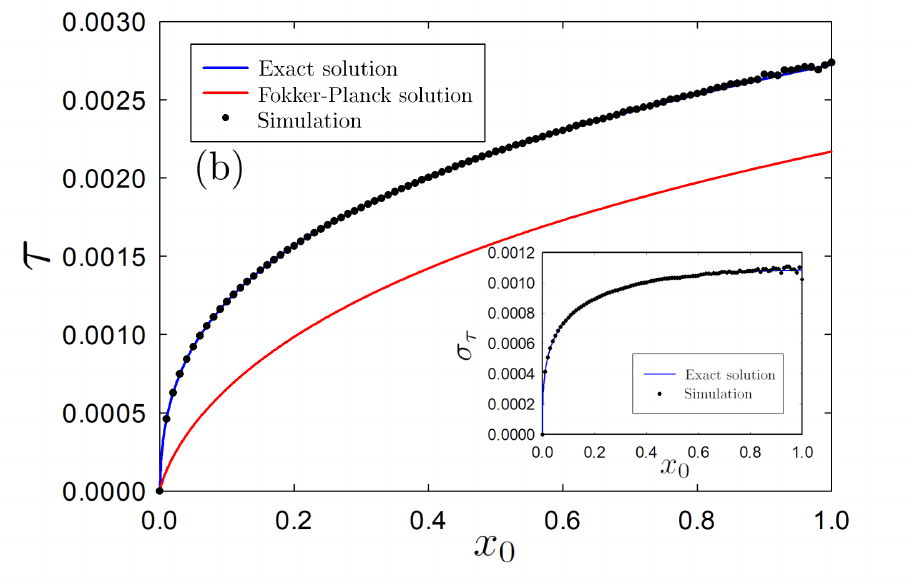}
    \caption{Mean persistence time for meme attention for both homogeneous (panel (a)) and scale-free (panel (b)) networks. The simulations (symbols) agrees perfectly with the exact solution (blue line) given by \eqref{eq:tau-exact} in both panels. Moreover, the diffusive approximation in \eqref{eq:back-FP} (red line) gives also a quite good estimation in the homogeneous network whereas that is not the case in the scale-free network. We have considered a system with $N=10^3$, an out-degree distribution $p_k=\delta_{k,10}$ and $\mu=15$ for the panel (a); while we have used $p_k \propto k^{-2.5}$ for $k \geq 4$ and $\mu=100$ for the panel (b). In the inset, the variance of the persistence time shows a perfect agreement with the exact solution \eqref{eq:tau2-exact}.}
   \label{Fig2}
\end{figure*}

\subsection{Diversity patterns in information ecosystems.} 

Biodiversity in natural ecological systems plays a fundamental role for preserving ecosystem functioning and health. Similarly, diversity of memes is a crucial aspect if we want to maintain a ``healthy'' information in social systems, where plurality of ideas circulate, in opposition to echo-chambered few polarized amplification of ideologies in the OSN.

In this section, we focus specifically on two different aspects of meme diversity: the number of different memes coexisting in the system and how user attention is distributed among such memes, i.e. how memes are distributed on users screens.

\subsubsection{Number of coexisting memes}

The capacity of attention of users in an OSN is finite and affect memes spreading \cite{gonccalves2011modeling,hodas2012visibility,qiu2017limited}. Therefore, a given network cannot sustain an arbitrarily large number of memes. Interestingly, due to both neutrality and constant innovation rate ($\mu$) assumption, it is possible to exactly calculate the probability $P_S$ of having $S$ different memes coexisting in the OSN \cite{Suweis_2012}. 
In fact, the stationary solution for $P_S$, reached in the large time limit, is a Poisson distribution
\begin{equation}\label{Poisson}
P_S = \frac{e^{- \left\langle S \right\rangle } \left\langle S \right\rangle^{S}}{S!}
\end{equation}
with an average number of memes given by
\begin{equation}
\label{eq:Smed}
\left\langle S \right\rangle = \tilde{\mu} \sum_{\nu_0=1}^N p_{\nu_0-1} \tilde{\tau}(\nu_0),
\end{equation}
that for the particular case of a delta peaked out-degree distribution $p_k=\delta_{k,k^*}$, simply reduces to $\left\langle S \right\rangle = \tilde{\mu}  \tilde{\tau}(k^* +1)$. Interestingly, the Poisson shape of the distribution of the memes diversity is ``universal'' regardless the details of the underlying network. The specific effect of the network structure is in determining the average number $\langle S \rangle$ of the coexisting memes in the systems.

This result was originally derived in the ecological context for a coarse-grain birth-death model for species diversity \cite{Suweis_2012}. However, in our model memes may appear with any $\nu_0$ initial number of users paying attention to it,  with a probability which is given by the out-degree distribution $p_{\nu_0-1}$. We can thus see how the diversity of ideas circulating in the network is promoted by large mean persistence times, and thus scale free networks are detrimental plurality of ideas in the OSN and promote strong polarization (and visibility) of few ideologies.

Note that, equivalently, we can formulate equation \eqref{eq:Smed} using the framework of continuous time. It simply reads $\left\langle S \right\rangle = N \mu \int_{0}^1 dx_0 \tau (x_0) p(x_0)$, where we have used that, in the continuous description, innovation is a Poisson process with rate $N \mu$ and $p(x_0)$ is the continuous counterpart of the out-degree distribution, giving the initial fraction of users attention when the meme appears in the OSN. Consistently, the fraction $\left\langle S \right\rangle /N $ depends only on the parameters of the continuous model.

\begin{figure*}
   \includegraphics[width=0.49\textwidth]{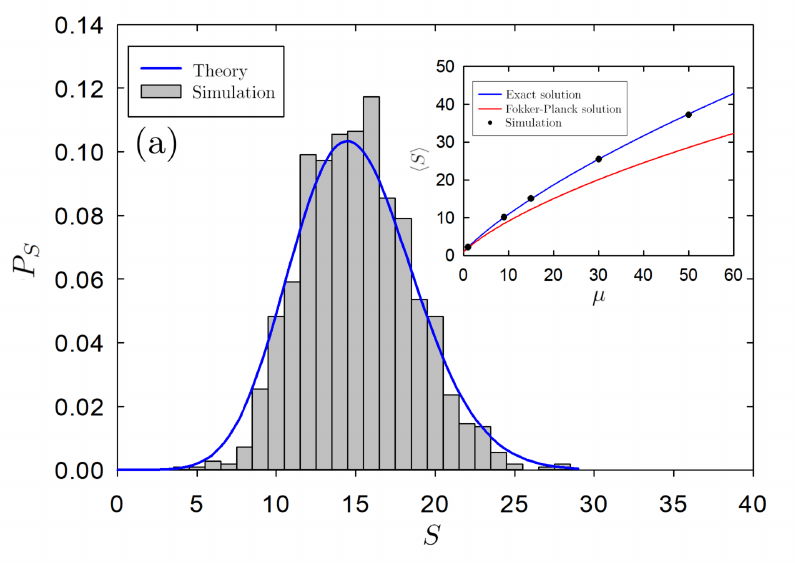}
   \includegraphics[width=0.49\textwidth]{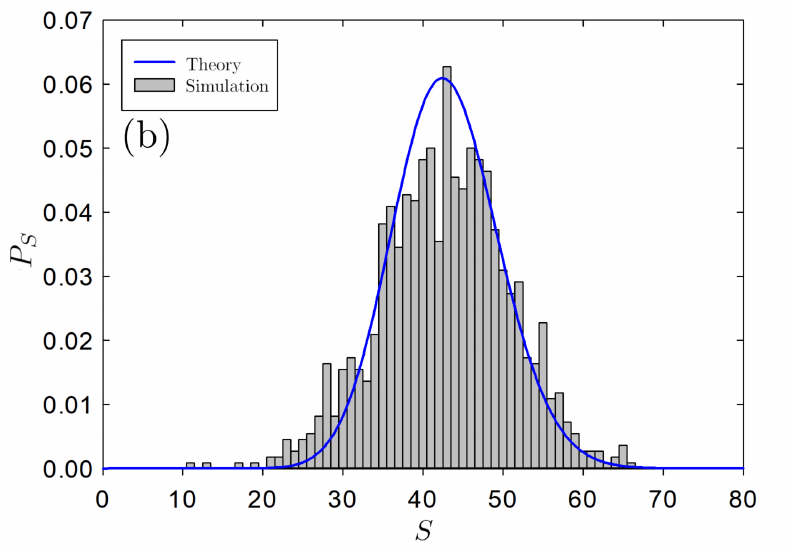}
    \caption{Histogram of number of different memes present in the whole system for both homogeneous (panel (a)) and scale-free (panel (b)) networks.  Numerical simulations agrees with the theoretically predicted Poisson distribution, especially for the homogeneous network. Note that there is no fitting parameter in the distribution since the mean is given by \eqref{eq:Smed}. We have considered the same model parameters to those chosen in  Fig.~\ref{Fig2}. Inset: Mean number of memes (species) $\left\langle S \right\rangle$ for different values of the innovation rate $\mu$. Apart from $\mu$, the rest of parameters are equal to those in the main figure.
 }  \label{Fig3}
\end{figure*}

In Fig.~\ref{Fig3} we have compared the theoretically predicted Poisson distribution with the histogram obtained in numerical simulations for long time in the systems studied above, that is, we consider $N=10^3$ and either $\mu=15$ and $p_k=\delta_{k,10}$; or  $\mu=100$ and $p_k \propto k^{-2.5}$ for $k \geq 4$. Moreover, in the homogeneous network, we have considered in the inset different values for the innovation rate $\mu$ to study the dependence  of the average number of memes. On the one hand, we obtain a perfect agreement when we use the exact calculation for $\tilde{\tau}$ as given by \eqref{eq:tau-exact} especially for the homogeneous network. Poisson distribution, explains quite successfully the pattern observed in the scale-free network although the quality of the agreement is slightly lower, but still quite good. On the other hand, although we did not expect a quantitative agreement (because of the bursty nature of the communication dynamics), also the prediction provided by the diffusive approximation \eqref{eq:tau_FP} remarkably shares the same qualitative behavior. As expected, both theories and simulations converge to $\left\langle S \right\rangle =1$ for the limit case $\mu \to 0$ since, in the absence of innovation, fluctuations eventually drive the system to the absorbing state and thus there is only one meme surviving at the end.

Finally, and motivated by the good agreement between theory and numerical simulations, we decided to test our predictions in a real scenario against data from the OSN Twitter. To do so, we collected samples of Twitter activity freely available from the Twitter Streaming API, spanning a period of $4$ years between $2015$ and $2018$, and comprising more than $135$ million tweets (see Appendix \ref{app:data} for details on the data collection). 

For the comparison, we assume that each meme in the model represents a hashtag in Twitter. Its popularity is traceable through the number of times it has been observed in the data set. However, given obvious privacy restrictions in the data that allow to see when a hashtag has been tweeted but not when it appeared on a screen, some quantities such as the attention paid to a hashtag at a given time cannot be directly measured. For these reasons we decided to limit the comparison only to observables that can be estimated directly from the data, such as the distribution $P_S$ of different memes present in the system and the mean number of co-existing memes $\langle S \rangle$ as a function of $\tilde{\mu}$ (see Appendix \ref{app:data} for the estimation of $\tilde{\mu}$ and $S$ from the data).

\begin{figure}
   \includegraphics[width=0.49\textwidth]{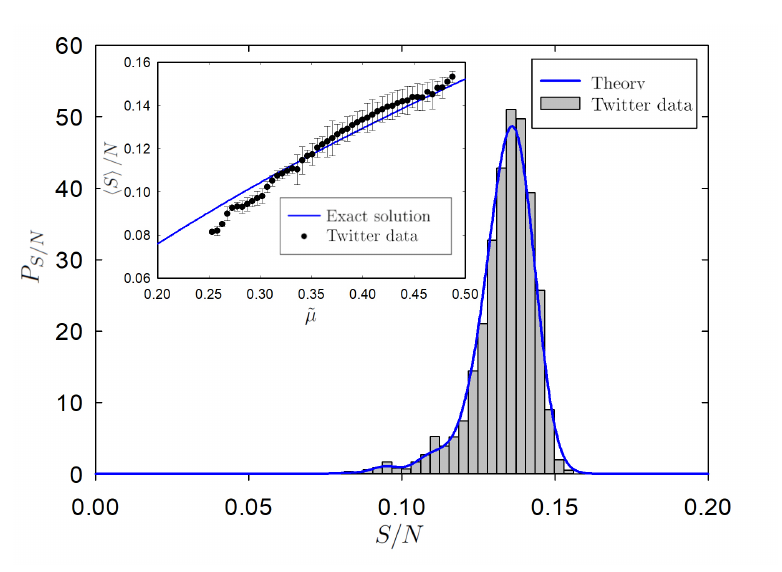}
    \caption{Comparison between theoretical predictions and Twitter data (without fit). Main panel: histogram of the normalized number of different memes present in the whole system obtained considering the full data set. For the Twitter data $S$ has been evaluated as the number of active hashtags at the end of each time slice. The theoretical distribution (blue line) has been calculated as a weighted average of Poisson distributions, each for value of $\tilde{\mu}$ observed in the data.  Inset: Mean number of memes (species) divided by the systems size $\left\langle S \right\rangle /N$ for different values of the innovation rate $\tilde{\mu}$. For the theoretical curve, we have assumed a power-law out-degree distribution with exponent 2.5 as the one presented in this work. }  \label{Fig5}
\end{figure}

In Fig.~\ref{Fig5} we show an analogous information to that presented in Fig.~\ref{Fig3}, but using the data extracted from Twitter. In this case, since each sample in the data has a different $\tilde{\mu}$ and different size, the theoretical distribution (blue line in  Fig.~\ref{Fig5}) has been obtained as a weighted average of different Poisson distributions given by Eq.~\eqref{Poisson}, one for each value of $\tilde{\mu}$ observed in the data (see Appendix \ref{app:data}). In the inset of Fig.~\ref{Fig5} instead, we repeated exactly the same analysis done for Fig.~\ref{Fig3}. Remarkably, given the correlations and possible biases in the data that could contradict the assumptions of our model, the good agreement  demonstrates that, at least at the macroscopic scale, such emergent human communication patterns in OSNs can be explained successfully by the proposed neutral model.   

\subsubsection{Relative Memes Attention}
       
Equations \eqref{Poisson} and  \eqref{eq:Smed} characterize diversity of memes at stationarity, but it gives no information on the relative attention that users give to such memes. We are thus interested in this new pattern that can be described by the probability $P_{RMA}$  that a meme receives attention from a fraction $x$ of users in the OSN. We name this pattern as relative memes attention (RMA), which is completely  analogous to the  relative species abundance  (RSA) pattern in ecology \cite{Volkov2003,vallade2003,azaele2016neutral}. This quantity is clearly relevant in communication dynamics, as it allows to understand how users' attention is distributed among memes. For example, when the RMA is peaked around a specific value, each meme attracts attention of a characteristic number of users. On the other hand, if RMA is fat tailed, then users' attention is strongly unevenly distributed among tweets, with some meme attracting attention of the majority of users.

Let us define $\phi(x)$ as the density function of the average number of memes receiving an attention $x$. Note that the only difference between $\phi(x)$ and $P_{RMA}(x)$ is the normalization constant, being the latter normalized to unity, while the former to the total number of different memes $S$. In the stationary state, the memes that contribute to the RMA are those which have been generated by innovation a time $t$ before and they are still present in the OSN. They thus contribute to $\phi(x)$ with an amount $\int_0^1 dx_0 p(x_0)P(x,t | x_0)$, which is integrated over all possible initial configurations $x_0$, weighted by the out-degree distribution. We have explicitly included the initial condition in the solution of the forward Fokker-Planck equation \eqref{eq:FP}. Since the number of species generated in a small time interval $dt$ is simply $\mu N dt$, the RMA finally reads \cite{vallade2003}
\begin{equation}
\label{eq:Vallade}
\phi (x) = \mu N \int_0^{\infty} dt  \int_0^1 dx_0 p(x_0)P(x,t | x_0).
\end{equation}

The application of Eq.~\eqref{eq:Vallade} is not straightforward, since we need the full time solution for $P(x,t|x_0)$. Although the Fokker-Planck equation \eqref{eq:FP} cannot be exactly solved,
nevertheless we have an approximated solution that can be obtained by neglecting the quadratic term in $x$ in the diffusion coefficient \eqref{eq:FP-coef2}, i.e. the diffusive term is simply $\propto x$ \cite{azaele2006}. This approximation is particularly good if one can assume that, during the time evolution, the invasion of most of the users screens by one single meme is unlikely. 

Plugging the full transient solution of $P(x,t|x_0)$ \cite{azaele2006} into \eqref{eq:Vallade} is neither immediate and an explicit analytical integration is not possible. However, if $x_0$ is small enough, a Taylor expansion of $P(x,t|x_0)$ in $x_0$ is suitable and makes the integration possible. Once the integral is carried out, we just need to take into account the normalization of $P_{RMA}$ to finally obtain
\begin{equation}
\label{eq:RMA}
P_{RMA}(x) = \frac{\exp\left( - 2 \frac{a}{b} x\right)}{x \psi\left( \frac{2a}{N b }\right)},
\end{equation}
where we recall that $\psi$  denotes the digamma function. The normalization has been imposed from $1/N$ that is the minimum non-vanishing value for attention fraction up to infinity. Note, that our assumptions legitimate the change of one by infinity in the upper bound. For a detailed derivation of \eqref{eq:RMA}, see Appendix \ref{app:RMA}.

We find that the outcome of our prediction for the RMA is a log-series distribution, rekindling a  classical result for RSA in ecological systems. \cite{Fisher1943}. Such RMA displays a power law behavior $p(x)\sim x^{-\beta}$ with an exponent $\beta=1$. Therefore, as expected, our model predicts that users' attention is strongly heterogeneously  distributed among memes, with most of them attracting a negligible number of users, while few (the viral ones) catching the attention of almost all of the OSN's users.

\begin{figure*}
   \includegraphics[width=0.49\textwidth]{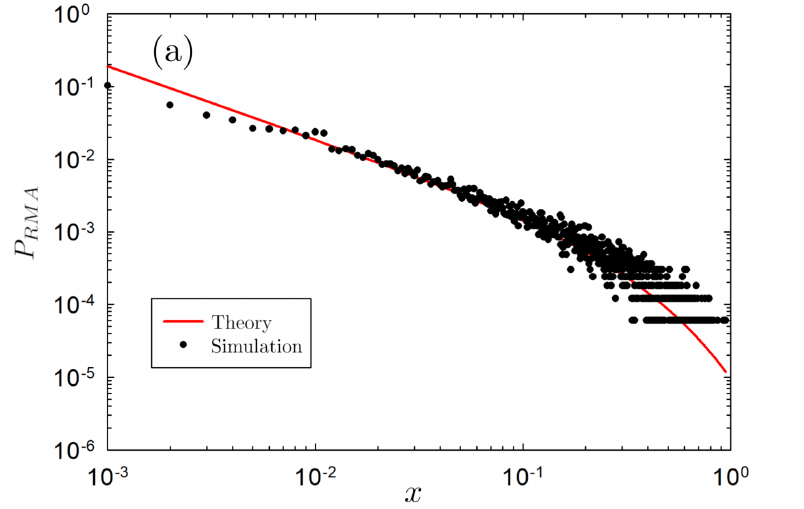}
   \includegraphics[width=0.49\textwidth]{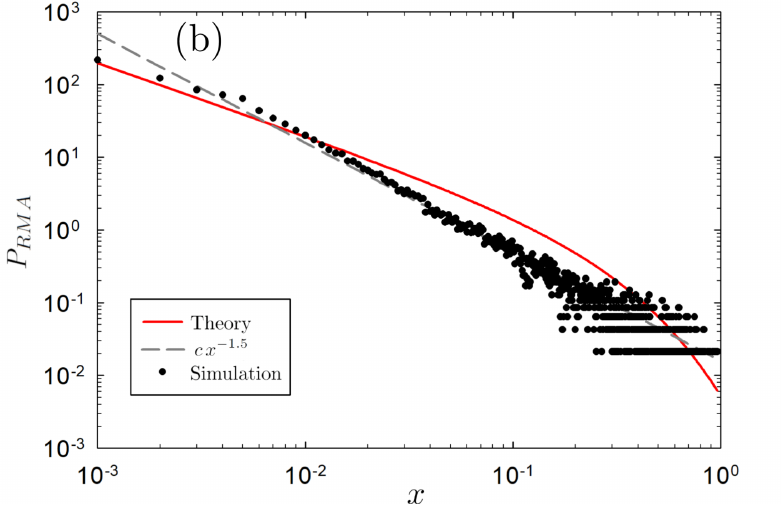}
    \caption{Relative memes attention for both homogeneous (panel (a)) and scale-free (panel (b)) networks. The probability $P_{RMA}(x)$  of finding a meme which receive a fraction $x$ of attention is plotted. Theoretical prediction (red line) given by \eqref{eq:RMA} reproduces remarkably well the tail of the pattern obtained by numerical simulations (symbols). We have considered the same model parameters to those chosen in  Fig.~\ref{Fig2}.}
   \label{Fig4}
\end{figure*}

We compare the log-series with the relative meme attention obtained from numerical simulations of the interacting particle model in Fig.~\ref{Fig4}. As before, the results for both networks homogeneous and scale-free are shown. On the one hand, in spite of the strong approximation we have carried out, the log-series reproduce very well the simulated pattern for the homogeneous network. On the other hand, as expected, we find that the approximated diffusive theory cannot reproduce such simulated pattern, which is the result of a bursty dynamics. Interestingly, in this latter case $P_{RMA}$ is compatible with a power-law  with exponent 1.5 in a broad range of attention regime. 


\section{Discussion and Conclusions}
\label{sec:conclusions}
Having high diversity in information ecosystem is not less important than having a rich biodiversity in natural ecosystems. In fact, diversity and heterogeneity of ideas in OSNs is a crucial aspect for the quality
of the deliberative process \cite{helberger2018challenging}. Online spaces dominated by one or few visions, i.e., with very low biodiversity, represent biased information ecosystems where phenomena such as fake news, filter bubbles, and echo chambers   crystallize
beliefs and annihilate
diverse opinions \cite{del2016echo}.
Therefore, being able to characterize diversity patterns in information ecosystems is not only a theoretically intriguing problem, but also an important aspect to measure their ``health''.

In this work, we have proposed an analytically tractable neutral theory to describe the dynamics of user attention to competing memes in OSNs. In particular, we have shown that we are able to  compute several new quantities of interest such as the number of coexisting memes, the distribution of user attention among these memes, and the average persistence time for attention on a meme. 
All these emergent properties have an ecological analogy in natural systems, suggesting that an ecological approach to study information ecosystems can open novel paths and understanding of the dynamics of memes in OSN.

By comparing our theoretical predictions to numerical simulations of the model, we have shown that the continuous approximation provides accurate results when the dynamics of the user attention $x$ is diffusive rather than bursty. This, in turn, depends on the underlying architecture of the user-user network. In fact, if the network is characterized by a scale-free degree distribution, then the diffusive approximation fails to quantitatively reproduce the biodiversity patterns, although the qualitative behavior is still well described. Using a backward semi-analytical Master Equation approach allows to overcome this limitation and to correctly predict the mean number of different coexisting memes and their mean persistence attention time. We note that having an analytical theory to calculate such quantities is of paramount importance, as it allows to easily understand the effect of the system parameters (such as the network connectivity or the innovation rate) to the persistence of active memes in the OSN and their diversity. Moreover, we note that, because of very strong finite size effects and fluctuations in the dynamics, simulations may lead to wrong conclusions especially when considering, as it typically is, large and heterogeneous systems.

Just as for the numerical simulations, the excellent agreement between the theory and the patterns extracted from the Twitter samples have important implications, both theoretical and practical. First of all, it indicates that our framework can provide information on the status of real information ecosystems. From few samples, it could be possible to estimate the plurality and ``health'' of online discourses.  Moreover, our results demonstrate that OSNs can be studied through the lens of neutral theory. Such a result draws a direct connection between information and natural ecosystem and allows the use of all the machinery developed in theoretical ecology
~\cite{azaele2016neutral} for the study of online human interactions. 

We would like to stress that the success of our theory in explaining and reproducing real patterns, does not translate in the claim that our model is capturing all the different mechanisms driving social interactions and meme spreading. Instead, our work show that we have put forward a ``simple enough'' model to explain and reproduce successfully some real emergent patterns in OSNs, hinting that for such properties, it is competition the dominating driver.  

Finally we note that, differently from popularity measures, user attentions to competing memes is not a feature that can be directly measured from data but can only be evaluated through some proxy. To this regard, a future perspective of this work is to connect mean persistence attention time and RMA to actual measurable proxies. Doing so, we could also understand if --and when-- the neutrality of the dynamics is broken. For instance, strong external events, such as breaking news, may have an impact to the attention dynamics that cannot be described in terms of demographic stochasticity, but calls for incorporating in this framework also environmental noise \cite{hidalgo2017species} and non-neutral effects \cite{martin2016eco}.

\begin{acknowledgments}
This project has been funded by Cariparo Foundation Visiting program 2019. SS and CP also acknowledge UNIPD Stats grant BioReact 2018. SM and VCS also acknowledge support from the Spanish Ministry of Science and Innovation, the AEI and FEDER (EU) under the grant PACSS (RTI2018-093732-B-C22) and the Maria de Maeztu program for Units of Excellence in R\&D (MDM-2017-0711). M.J.P, A.S-R. and J.B-H. acknowledge the support of the Spanish MICINN project PGC2018-096999-A-I00. M.J.P. acknowledges as well the support of a doctoral grant from the Universitat Oberta de Catalunya (UOC).
\end{acknowledgments}

\appendix


\section{Transition probabilities and diffusive approximation}
\label{app:FP}

We put forward explicitly, in this appendix, the functional form of the transition probabilities assumed in our model of OSN. As explained in the main text, there are three different classes of transformations that correspond to:
\begin{itemize}
\item  Spread of a node that holds the meme $A$ and has $k$ followers, $j$ of the which hold the meme $B$. The probability of such a transition in a system with $\nu$ nodes carrying meme $A$ is
\begin{equation}
\label{eq:tran-As}
W^{A,s}_{k,j}(\nu)= \frac{\nu}{N} \left( 1-\tilde{\mu} \right) p_k \dfrac{\dbinom{\nu-1}{k-j}\dbinom{N-\nu}{j}}{\dbinom{N-1}{k}}.
\end{equation}
\item Innovation of a node that holds the meme $A$ and has $k$ followers, $j$ of the which hold the meme $A$. The probability of such a transition in a system with $\nu$ nodes carrying meme $A$ is
\begin{equation}
\label{eq:tran-Ai}
W^{A,i}_{k,j}(\nu)= \frac{\nu}{N} \tilde{\mu} p_k  \dfrac{\dbinom{\nu-1}{j}\dbinom{N-\nu}{k-j}}{\dbinom{N-1}{k}} .
\end{equation}
\item Action of a node that holds the meme $B$ and has $k$ followers, $j$ of the which hold the meme $A$. The probability of such a transition in a system with $\nu$ nodes carrying meme $A$ is
\begin{equation}
\label{eq:tran-B}
W^{B}_{k,j}(\nu)= \frac{N - \nu}{N}  p_k \dfrac{\dbinom{\nu}{j}\dbinom{N-\nu-1}{k-j}}{\dbinom{N-1}{k}}.
\end{equation}
\end{itemize}
All expressions are obtained as the product of four probability factors. The first one is the probability of choosing the node that acts. The second one is the probability of the type of event either spread or innovation. The third one is the probability of having $k$ followers.  The last one is the hypergeometric probability coming from the distribution of the possible states of the $k$ followers. Note that the ``second'' factor in \eqref{eq:tran-B} is a factor 1 that is not explicitly written. 

The Master Equation \eqref{eq:ME} is simply a balance of gained and lost probability due to transitions between states. Note that  $\nu - \Delta \nu$ appears as the argument of $P$   with $\Delta \nu$ being the change in the number of users paying attention to node $A$.

In order to perform the diffusive approximation, we consider the limit $N \to \infty$, with $x=\nu/N$. In such a limit, we can approximate the hypergeometric distribution with a binomial one, which yields the continuous description of the transition rates
\begin{subequations}
\label{eq:W-cont}
\begin{align}
W^{A,s}_{k,j}(x)&= x \left( 1-\tilde{\mu} \right) p_k  x^{j}(1-x)^{k-j},\\
W^{A,i}_{k,j}(x)&= x \tilde{\mu} p_k  (1-x)^{j}x^{k-j},
\\
W^{B}_{k,j}(x)&= (1-x)  p_k x^j (1-x)^{k-j}.
\end{align}
\end{subequations}
Carrying out an expansion of the Master Equation \eqref{eq:ME} in powers of $N^{-1}$, introducing transition probabilities as in \eqref{eq:W-cont} and assuming scaling of time $t=\tilde{t}/N^2$ and innovation rate $\mu=N\tilde{\mu}$ as introduced in the main text, we achieve the diffusive approximation given by the Fokker-Planck equation \eqref{eq:FP} with the coefficient reported in \eqref{eq:FP-coef}.

\section{Backward Fokker-Planck equation and lifetime distribution}
\label{app:FP-back} 

We have obtained the forward Fokker-Planck equation
\begin{equation}
\partial_t P(x,t) = \partial_x \left[a x P(x,t) \right] + \frac{1}{2} \partial_x^2 \left[ bx(1-x) P(x,t)\right].
 \end{equation}
The backward version of this equation,
\begin{equation}
\label{eq:ap-FP-b}
\partial_t Q(x_0,t) = -a x_0 \partial_{x_0}  Q(x_0,t)  + \frac{bx_0(1-x_0)}{2} \partial_{x_0}^2   Q(x_0,t),
\end{equation}
governs the probability $Q(x_0,t)$ of reaching the absorbing boundary placed at $x=0$ departing from $x_0$ after an evolution of time $t$. Consequently, the differential equation that holds for the probability of reaching the absorbing point at any time $\Pi(x_0)=\int_0^{\infty}dt\, Q(x_0,t)$ is 
\begin{equation}
0 =- a x_0 \partial_{x_0}  \Pi(x_0)  + \frac{bx_0(1-x_0)}{2} \partial_{x_0}^2   \Pi(x_0)
\end{equation}
where we have used that $Q(x_0,0)=\lim_{t\to\infty}Q(x_0,t)=0$. Since, the only absorbing boundary is that at $x=0$, the meaningful solution for the previous equation is $\Pi(x_0)=1$. In other words, all the memes eventually will reach the extinction.

To study the mean lifetime $\tau(x_0)$ of the memes we need to multiply the backward equation \eqref{eq:ap-FP-b} by $t$ and then integrate for all time. After assuming regular behavior for the boundary terms and taking into account that $\Pi(x_0)=1$, we get
\begin{equation}
1=- a x_0 \partial_{x_0}  \tau(x_0)  + \frac{bx_0(1-x_0)}{2} \partial_{x_0}^2   \tau(x_0).
\end{equation} 
Imposing the boundary conditions $\tau(0)=0$ and $\lim_{x_0 \to 1} \tau(x_0) <\infty $ we obtain solution \eqref{eq:tau_FP} presented in the main text.

\section{Backward Master Equation and lifetime distribution}
\label{app:ME-back}
The backward Master Equation that rules the dynamics of the probability $Q(\nu_0,\tilde{t})$ of reaching extinction departing from a initial attention of $\nu_0$ after $\tilde{t}$ time steps is written in \eqref{eq:back-ME}. The probability in the next time step  is a linear combination of the probability in the current one. Therefore, we can write a vector equation of dimension $N+1$ using matrix notation
\begin{equation}
\widehat{\bm{Q}}(\tilde{t}+1)= \widehat{\mathbb{M}} \widehat{\bm{Q}}( \tilde{t})
\end{equation} 
with $\widehat{\bm{Q}}(\tilde{t})$ being the vector with components $Q(\nu_0,\tilde{t})$, for $\nu_0=\{0,\ldots,N\}$,
and $\widehat{\mathbb{M}}$ the $(N+1)\times(N+1)$ matrix with elements
\begin{align}
M_{\nu_0,\nu_0'}= \sum_{k=0}^{N-1} \bigg[ &W_{k,\nu_0'-\nu_0}^{A,s}(\nu_0) + W_{k,\nu_0-\nu_0'-1}^{A,i}(\nu_0)\nonumber \\ & +W_{k,\nu_0-\nu_0'}^{B}(\nu_0) \bigg].
\end{align}
In the previous equation, to prevent from clutter our formulae, we assume that transition probabilities are equal to zero when the indexes involves a transformation with no physical meaning, e.g.~second subindex does not belong to the interval $[0,k]$. Probability conservation is reflected in the property
\begin{equation}
\label{eq:norm-M}
\sum_{\nu_0'=0}^{N}M_{\nu_0,\nu_0'} =1.
\end{equation}

The absorbing boundary condition at $\nu=0$ makes that $Q(0,\tilde{t})=\delta_{\tilde{t},0}$. Hence, in dimension $N$, we have that
\begin{equation}
\label{eq:BME-N}
\bm{Q}(\tilde{t}+1)= \mathbb{M} \bm{Q}( \tilde{t})+ \bm{M}_0 \delta_{\tilde{t},0}
\end{equation} 
where $\bm{Q}(\tilde{t})$ is the vector $\widehat{\bm{Q}}(\tilde{t})$ after removing the first component corresponding to $\nu_0=0$, $\mathbb{M}$ is the submatrix of $\widehat{\mathbb{M}}$ obtained after removing the rows and columns corresponding to value $0$, and $\bm{M}_0$ is a vector column that contains the transition probabilities to the extinction state.

We define the total probability of arriving 0 departing from $n$ as the time sum
\begin{equation}
\bm{\Pi}=\sum_{\tilde{t}=0}^\infty  \bm{q}(\tilde{t}),
\end{equation}
where the components of $\bm{\Pi}$ are $\Pi(\nu_0)$.
Summing the backward Master Equation \eqref{eq:BME-N} for all time, and taking into account that $\bm{Q}(0)=\lim_{\tilde{t}\to\infty}\bm{Q}(\tilde{t})=0$, we get
\begin{equation}
\label{eq:Pexit}
\bm{\Pi}=\mathbb{M} \bm{\Pi}+ \bm{M}_0 .
\end{equation}
Since property \eqref{eq:norm-M} holds, we have that the solution for the previous equation is simply $\bm{\Pi}=\bm{1}$, where we have used the same notation for the vector of dimension $N$ full of ones that in the main text. Consistently with our results in the continuous description and with our physical understanding, we get, in this exact framework, that all the memes will reach extinction eventually. 

The mean first passage time $\tilde{\tau}(\nu_0)$ to reach the absorbing point starting from the state $\nu_0$ is given by
\begin{equation}
\tilde{\bm{\tau}} = \sum_{\tilde{t}=0}^\infty  \tilde{t} \, \bm{Q}(\tilde{t}).
\end{equation}
Multiplying \eqref{eq:BME-N} by $\tilde{t}$ and summing for all time, we obtain the equation for the mean persistence time \eqref{eq:tau-exact0} presented in the main text with its corresponding solution \eqref{eq:tau-exact}.

It is also possible to compute the average of the square passage time, which is useful for the variance. We define
\begin{equation}
\left\langle \bm{\tilde{t}}^{2} \right\rangle = \sum_{\tilde{t}=0}^\infty  \tilde{t}^2  \bm{Q}(\tilde{t}).
\end{equation}
Multiplying \eqref{eq:BME-N} by $\tilde{t}^2$ and summing for all time, we obtain
\begin{equation}
\left\langle \bm{\tilde{t}}^{2} \right\rangle  -2\tilde{\bm{ \tau}}+\bm{1} = \mathbb{M} \left\langle \bm{\tilde{t}}^{2} \right\rangle,
\end{equation} 
which has the solution \eqref{eq:tau2-exact} presented in the main text.

\section{Derivation of the relative meme attention }
\label{app:RMA}
Our starting point here is the approximated Fokker-Planck equation after neglecting the squared term in the diffusion coefficient, that is,
\begin{equation}
\label{eq:eq-S}
\partial_t P(x,t|x_0) = \partial_x \left[a x P(x,t|x_0) \right] + \frac{1}{2} \partial_x^2 \left[ bx P(x,t|x_0)\right].
 \end{equation}
From now on, we consider that the equation is defined in the region $x>0$. We understand the neglect of the squared term as a rescaling of the attention that moves the boundary from $x=1$ to infinity. The solution of \eqref{eq:eq-S} submitted to an absorbing boundary at $x=0$ and the initial condition $P(x,t|x_0)=\delta(x-x_0)$ is \cite{azaele2006}
\begin{align}
P(x,t|x_0)=&\frac{2a}{b}\frac{1}{1-e^{-at}} \exp\left(- \frac{2a}{b} \frac{x+x_0 e^{-at}}{1-e^{-at}} \right) \nonumber \\ &\times \left(\frac{x_0}{x} e^{-at}\right)^{\frac{1}{2}} I_1 \left( \frac{4a}{b} \frac{\sqrt{x_0 x e^{-at}}}{1-e^{-at}}\right).
\end{align}
A Taylor expansion of this solution up to linear order yields
\begin{equation}
P(x,t|x_0) \simeq \frac{4a^2 \exp{\left[-\left( at + \frac{2a}{b} \frac{x}{1-e^{-at}}\right) \right]}}{b^2 \left( 1- e^{-at}\right)^2}x_0.
\end{equation}
With this approximation, valid for small $x_0$, we can carry out exactly the integration for all time that results
\begin{equation}
\label{eq:ap-D4}
\int_0^{\infty} dt  P(x,t|x_0) \propto \frac{\exp \left(-2\frac{a}{b}x \right)}{x},
\end{equation}
where we have explicitly written the dependence with $x$. Since the relative meme attention $P_{RMA}$ is proportional to the integral in \eqref{eq:ap-D4}, we just need to impose the normalization condition. Taking into account that the minimum non-vanishing value for the attention is $1/N$, and assuming, as said above, that the upper bound is at infinity, we end up with the $P_{RMA}$ reported in equation \eqref{eq:RMA} of the main text.

\section{Simulation details}
\label{app:sim}
In this appendix, we cover in detail the algorithm used to carry out the numerical simulations presented in this work. We have explicitly simulated the microscopic dynamics described in the main text. Specifically, we consider a system made by $N$ nodes. Each of them carries just one meme. The dynamics is generated by repetition of the next recipe for each time step:
\begin{itemize}
\item We randomly choose a node with homogeneous probability $1/N$.
\item With probability $\tilde{\mu}$, the node innovates the meme.
\item The degree of the node $k$ is chosen from the out degree distribution $p_k$.
\item From the available $N-1$ nodes, we randomly choose the $k$ followers.
\item The meme of the firstly chosen node spreads to all the followers.
\end{itemize}  
Note that the connections in the network are not fixed, we work with a random network which is dynamic. This random dynamic feature guarantees that the approach given by the Master Equation used in the main text is the correct mathematical description of our system.

For each set of parameters, we have run $N_S=100$ simulations of a total duration $\tilde{t}_f = 5 \cdot 10^6$ time steps. The measurements that correspond to the persistence time comes from the average of all registered times for those memes which reach extinction inside the simulation window. Recording the history of all memes, it is easy to recover the times to extinction starting from all possible initial attentions. We just need to  subtract the time in which the meme has that initial attention to the extinction time of the corresponding meme. We have increased the statistics of the measurements of biodiversity, which are performed in the stationary state, taking different times of observation. We have taken not only $\tilde{t}_f$, but also $\tilde{t}_f - n \, \Delta \tilde{t}$ with $n=1,\ldots,10$ and $\Delta \tilde{t} = 2.5 \cdot 10^5$ time steps. The choice of $\Delta t$ has to guarantee some requirements. On the one hand, $\Delta \tilde{t}$ has to be big enough to prevent correlation between the different observation times. On the other hand, $\Delta \tilde{t}$  has to be small enough to assure that the system is close enough to the stationary state in the first observation time $\tilde{t}_f - 10 \Delta\tilde{t}$. 

In general, in the main text we report all values using the continuous scales, which is related with the discrete dynamics through a factor $N^2$ for the timescale, $t=\tilde{t}/N^2$, and a factor $N$ for the attention $x=\nu /N$.

\section{Twitter data}
\label{app:data}
In this appendix, we describe the details of the data collection from the OSN Twitter and the analyses used in the comparison with the theoretical predictions. In order to get an unbiased sample of Twitter activity, we collected data from the Twitter Streaming API. The data collection covers a period of $4$ years from January $2015$ to December $2018$. Only geolocalized tweets originated in a rectangle area covering the UK have been requested. This latter condition allowed us to avoid the $1\%$ total traffic limitations imposed by the Streaming API and to assure that the majority of tweets and hashtags would be in English. Finally, to protect privacy, users' information and the tweetID have been anonymized through a hash function before their storage and, in any case, for each tweet, only the timestamp and the hashtags contained in the text have been used in the analyses. The complete data set comprises approximately $135$ million tweets containing, at least, one hashtag.

To reduce the possible noise generated by misspelled hashtags or bots, we  filtered out all the hashtags that have been twitted less than --i.e. have a minimum popularity of--  $5$ times. We have explicitly checked that our results remain almost unchanged under different filters between $3$ and $500$. After the filtering and with the aim of different samples for the statistical analyses, we divided the final data set in $10^{4}$ bins/samples, each of the size of around $12000$ tweets. Moreover, to test the robustness of our results with respect to the choice of the number and size of the bins, we also repeated the calculations for $10^{3}$, $2\times 10^{4}$ and $5\times 10^{4}$ bins along with one test in which we considered one day in the data as one bin. In all the cases, we find robust results that behave qualitatively in the same way that to those presented in Fig.~\ref{Fig5}.

Once we filtered and divided the data, for each bin/sample, we estimate the innovation rate $\tilde{\mu}$ as the fraction of new hashtags --i.e. used for the first time-- observed in a bin over the total number of unique hashtags observed until that bin. Note that, therefore, the innovation rate does not remain constant. This dynamics is in contrast to our model. Nevertheless, each bin/sample can be thought of as a realization of our model with different innovation rates.    For $S$/N, we take the number of unique hashtags surviving at the end of each bin normalized by the maximum value of $\langle S \rangle$/N given by the theory. The histogram in Fig.~\ref{Fig5} is made just binning and counting these data along the $S$/N axis. In order to get a better statistics to determine average values, we group the samples binning in $\tilde{\mu}$, where we call $n_{\tilde{\mu}}$ to the number of samples considered within the bin with center at $\tilde{\mu}$. Therefore, we obtain in this way an averaged number of coexistent memes $\langle S \rangle /N$ and its variance $\sigma_{S/N}^2$ for each considered innovation rate. The inset of Fig.~\ref{Fig5} is constructed with these averages, where the size of the errorbars are given by the $\sigma_{S/N}$.

For the construction of the theoretical prediction in Fig.~\ref{Fig5}, we have used, in an auxiliary way the full data set, that is, the full list of corresponding $(\tilde{\mu},n_{\tilde{\mu}},\langle S \rangle/N)$. Specifically, we have used a weighted combination of Poisson distributions, one for each bin of $\tilde{\mu}$ with its observed $\langle S \rangle /N$. In other words, we have defined a weight $w_{\tilde{\mu}}=n_{\tilde{\mu}}/\sum_{\tilde{\mu}'} n_{\tilde{\mu}'}$ and we have built the distribution
\begin{equation}
  P_S=   \sum_{\tilde{\mu}}w_{\tilde{\mu}} P^{(\tilde{\mu})}_S
\end{equation}
as the linear combination of Poisson distribution $P^{(\tilde{\mu})}_S$ parameterized by the average $\langle S \rangle$ corresponding to that $\tilde{\mu}$. 
Note that, in order to combine the Poisson distributions, and then rescale from $S$ to $S/N$, the value of $N$ is required. Therefore, we have exploited the Poisson theoretical prediction associating an effective size to the data set given by 
\begin{equation}
N=\frac{\langle S \rangle /N}{\sigma_{S/N}^2},
\end{equation} 
which is also averaged over bins.

For the theoretical prediction in the inset, since Twitter users' out-degree (followers) distribution is extremely fat-tailed~\cite{gleeson2016effects,twitterpowerlaw}, in the model we assumed the same power-law degree distribution than the shown in the main text. We have used such parameters, also introduced in \cite{gleeson2014competition}, in the absence of a better knowledge of the real network. 

\bibliography{biblo_neutral_twitter}

\end{document}